\documentclass{article}[12pt] 
\usepackage{amssymb} 
\usepackage{amsmath}
\usepackage{amsthm}
\usepackage{amsfonts}
 \usepackage{times}
\usepackage[matrix,arrow]{xy}
\usepackage{tikz}
\usetikzlibrary{shapes, arrows, shadows}

\def\be{\begin{equation}}
\def\ee{\end{equation}}
\def\bea{\begin{eqnarray}}
\def\eea{\end{eqnarray}}

\def\thefootnote{\fnsymbol{footnote}}
\begin{document}
\begin{titlepage}
March 6, 2013         \hfill
\begin{center}

\vskip .5in
\renewcommand{\thefootnote}{\fnsymbol{footnote}}
{\Large \bf  Renormalization group defects for boundary flows    }\\
 
\vskip .50in

\vskip .5in
{\large Anatoly Konechny}\footnote{email address: anatolyk@ma.hw.ac.uk}

\vskip 0.5cm
{\large \em Department of Mathematics,\\
Heriot-Watt University,\\
Riccarton, Edinburgh, EH14 4AS, UK \\
and\\
Maxwell Institute for Mathematical Sciences\\
Edinburgh, UK}
\end{center}

\vskip .5in

\begin{abstract} \large
 Recently Gaiotto \cite{Gaiotto}  considered conformal defects which produce an expansion of   
 infrared local fields in terms of the ultraviolet ones for a given renormalization group flow.
 In this paper we propose that for a boundary RG flow in two dimensions there exist 
 boundary condition changing fields (RG defect fields) linking   the UV and the IR conformal boundary 
 conditions which carry similar information on the expansion of boundary fields at the fixed points. 
 We propose an expression for  a pairing between IR and UV operators in terms of a four-point function 
 with two insertions of the RG defect fields. For the boundary flows in minimal models triggered by $\psi_{13}$ perturbation 
 we make an explicit proposal for the RG defect fields. We check our conjecture by a number of calculations done for the example of 
  $(p,2)\to (p-1,1)\oplus (p+1,1)$   flows.

\end{abstract}
\end{titlepage}
\large

\newpage
\renewcommand{\thepage}{\arabic{page}}
\setcounter{page}{1}
\setcounter{footnote}{0}
\renewcommand{\thefootnote}{\arabic{footnote}}
\large
\section{Introduction}
\renewcommand{\theequation}{\arabic{section}.\arabic{equation}}

In perturbative renormalization the operators of perturbed theory are expressed in terms of the operators in the unperturbed theory with counter terms provided. 
At the level of the deformed OPE algebra such expressions are free from infrared divergences and are essentially perturbative under a very broad set of assumptions \cite{GuidaMagnoli}. 
The non-perturbative data needed 
to calculate correlation functions can 
be put into vacuum expectation values (see e.g. \cite{AleshaZam}, \cite{GuidaMagnoli}, \cite{Litvinovetal}).  One can imagine that such expressions hold all the way into the new infrared (IR) fixed point and that one 
can express the operators $\phi_{i}^{\rm IR}$ in the infrared fixed point in terms of linear combinations of operators in the UV fixed point:
\be \label{mapping}
\phi_{i}^{\rm IR} = \sum_{j} b_{ij} \phi_{j}^{\rm UV} \, . 
\ee  

When the two fixed points are near in the parameter space one can calculate the expansion coefficients $b_{ij}$  perturbatively. 
This is the case for RG flows between neighbouring minimal models $M_{m}$ and $M_{m-1}$ first considered in \cite{Zamolodchikov_pert}.    

The idea to associate a conformal defect (or domain wall) with a renormalization group (RG) flow was first put forward in \cite{BR} (see also section 5 of \cite{FQ})  
in the context of two-dimensional quantum field theories. We refer the reader to 
\cite{PetkovaZuber}, \cite{Bachasetal}, \cite{GrahamWatts}, \cite{Frohlichetal}, \cite{BachasGaberdiel} for a definition and discussion  of  general conformal 
defects. 
Recently a more concrete proposal relating defects to (\ref{mapping}) was made in \cite{Gaiotto}. It was proposed    that given a flow between two two-dimensional conformal field theories (CFT's) there exists a conformal defect 
between  the fixed point CFT's which allows one to calculate the coefficients $b_{ij}$.   The prescription of \cite{Gaiotto} is as follows. Assume the fields $\phi_{j}^{\rm UV}$ are canonically normalized.  Let $|\rm{RG}\rangle$ be the conformal boundary state in the tensor product ${\rm CFT}_{\rm UV}\otimes {\rm CFT}_{\rm IR}$ 
that represents the RG defect via the folding trick (see e.g. \cite{Bachasetal}), then 
\be\label{pairing1}
b_{ij}= \langle \bar \phi_{j}^{\rm IR}\otimes \phi_{i}^{\rm UV}|{\rm RG}\rangle 
\ee
where $\bar \phi_{j}^{\rm IR}$ are  the fields reflected by the folding (this essentially exchanges the holomorphic and antiholomorphic components). 
Alternatively, representing the RG defect by an operator $\widehat{\rm RG}:{\cal H}_{\rm UV} \to {\cal H}_{\rm IR}$ we have 
\be\label{pairing2}
b_{ij}=\langle \phi_{j}^{\rm IR}| \widehat{\rm RG}|\phi_{i}^{\rm UV} \rangle\, . 
\ee  
The last formula corresponds to putting the RG defect on the unit circle and computing a correlator with  $ \phi_{i}^{\rm UV} $ inserted at the origin and $\phi_{j}^{\rm IR}$ at infinity. 
\vspace{5mm}

\begin{center}
\begin{tikzpicture}
\filldraw[ultra thick, fill= gray!40!white, draw=black] (0,0) circle (1.5);
\draw (0,0) node {$\bullet$} ;
\draw (-0.2,0.2) node {$\phi^{\rm UV}_{i}$};
\draw (5,0) node {$ \phi_{j}^{\rm IR}(\infty)$};
\end{tikzpicture}
\vspace{5mm}

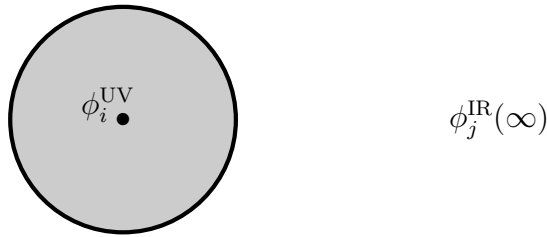
Fig1: Bulk RG defect.
\end{center}

In \cite{Gaiotto} an explicit algebraic construction was put forward for the RG defect corresponding to the flows between two neighbouring minimal models. 
It was shown that the leading order mixing coefficients calculated in \cite{Zamolodchikov_pert} are reproduced precisely by the proposed RG defect via (\ref{pairing1}).  
 
The idea that such RG defects exist in general seems very attractive. The complicated data of the mapping of  fields (\ref{mapping}) done by the RG flow can be encoded algebraically 
in the defect boundary field which can be accessed using  the techniques of conformal field theory. If one understands better the general properties of such RG 
defects this could lead to selection rules for  possible RG flows between known 2d CFT's.

Heuristically the existence of such RG defects can be argued for using the following construction. Consider a perturbed CFT. Put the perturbed theory on a plane with a non-trivial metric profile in the radial direction: 
$dx^{2} =  g^{2}(r)(dr)^2 + r^{2}(d\theta)^2$. Let the scale function $g(r)$ interpolate between a scale $\Lambda_{\rm UV}$ near the origin and  the scale $\Lambda_{\rm IR}$ at infinity. An operator $\phi^{\rm IR}$ defined 
at scale $\Lambda_{\rm IR}$ can be transported to an operator at scale $\Lambda_{\rm UV}$ by simply moving  it towards the origin. One can then imagine a limiting process for which the region in which the scale function $g(r)$
changes  shrinks to a small neighbourhood of $r=1$ and the constant scales $\Lambda_{\rm UV}$ , $\Lambda_{\rm IR}$ are sent to 0 and $\infty$ respectively. We obtain a domain wall between the UV and IR fixed points. 
 
Another possible general construction of the RG defect proceeds by perturbing the UV theory on a half plane and letting it flow with the RG \cite{BR}. 
As discussed in \cite{Gaiotto}  renormalizing such perturbations may require switching on new fields localized on the edge of the half plane the interpretation of which is unclear.

Both of these heuristic  constructions may be extended to the case of the boundary flows in which the bulk CFT is kept fixed and the flow interpolates between two 
conformal boundary conditions in this bulk theory. Naturally such defects would be point-like and thus must be represented by some boundary condition changing fields: 
$\psi^{\rm UV, IR}$ (with the conjugate counterpart  $\psi^{\rm IR,UV}$).  
By analogy with the bulk picture of Fig. 1 we may expect the RG paring for boundary fields to be given by a four-point function
\be 
\langle \psi^{\rm IR}_{j}(\infty) \psi^{\rm IR, UV}(-x)\psi^{\rm UV}_{i}(0)\psi^{\rm UV,IR}(x)\rangle \, ,
\ee 
see Fig. 2.
\vspace{1cm}

\begin{center}
\begin{tikzpicture}
\draw[very thick] (0,0)--(3,0);
\draw (3,0) node {x};
\draw[very thick, dashed] (3,0) -- (7,0);
\draw (5,0) node {$\bullet$};
\draw (7,0) node {x};
\draw[very thick] (7,0) -- (10,0);
\draw (3.1,0.5) node {$\psi^{\rm IR,UV}$};
\draw (5,0.5) node {$\psi_{i}^{\rm UV}$};
\draw (7.3,0.5) node {$\psi^{\rm UV,IR}$};
\draw (12,0) node {$\psi_{j}^{\rm IR}(\infty)$};
\end{tikzpicture}
\vspace{6mm}

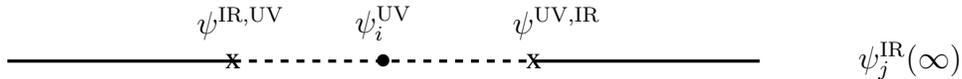
Fig2: RG pairing by boundary RG defects. The defect field insertions are marked by crosses.
\end{center}

\vspace{1cm}

Assuming the fields $\psi^{\rm UV, IR}$, $\psi^{\rm IR, UV}$ are   quasi primary of dimension $\Delta$ and the fields $ \psi^{\rm IR}_{j}$,   $ \psi^{\rm UV}_{i}$ 
are quasi primaries of dimensions $\Delta_{\rm IR}$ and $\Delta_{\rm UV}$ respectively one has 
\bea 
&& \langle \psi^{\rm IR}_{j}(\infty) \psi^{\rm IR, UV}(-x)\psi^{\rm UV}_{i}(0)\psi^{\rm UV,IR}(x)\rangle = \nonumber \\
&&\Bigl[ \frac{x^{\Delta_{\rm IR}-\Delta_{\rm UV}-2\Delta}}{4^{\Delta}}   \Bigr]
\langle  \psi^{\rm UV,IR}(\infty)\psi^{\rm IR}_{j}(1) \psi^{\rm IR,UV} \left(\frac{1}{2}\right)\psi^{\rm UV}_{i}(0)\rangle \, . 
\eea
Stripping off the prefactor we propose an analogue of the RG pairing (\ref{pairing1}), (\ref{pairing2}) for  the boundary flows to be given by 
\be \label{prescription1}
\langle \psi^{\rm IR}_{j}, \psi^{\rm UV}_{i}\rangle_{\rm RG} = \langle  \psi^{\rm UV,IR}(\infty)\psi^{\rm IR}_{j}(1) \psi^{\rm IR,UV} \left(\frac{1}{2}\right)\psi^{\rm UV}_{i}(0)\rangle
\ee
so that for canonically normalized basis $\psi_{j}^{\rm UV}$
\be \label{prescription2}
\phi_{i}^{\rm IR} = \sum_{j} b_{ij} \phi_{j}^{\rm UV}\, , \qquad b_{ij} = \langle \psi^{\rm IR}_{j}, \psi^{\rm UV}_{i}\rangle_{\rm RG} \, .
\ee

In the rest of the paper we focus on flows between conformal boundary conditions in minimal A-type unitary models which are triggered by the boundary field $\psi_{(1,3)}$. 
It was shown in \cite{RRS} that if one starts with a single Cardy boundary state  with labels $(a_1,a_2)$ and switches on $\psi_{(1,3)}$ the end point of the flow 
is the following superposition of Cardy boundary conditions
\be \label{flows}
(a_1,a_2)\longrightarrow \bigoplus_{i=1}^{{\rm min}(a_1,a_2)}(a_1+a_2 +1 -2i,1)\, . 
\ee
Since the IR boundary condition has $N={\rm min}(a_1,a_2)$ components the RG defect field $\psi^{\rm IR,UV}$ 
breaks up into  $N$ components as well. We propose that up to normalization the corresponding components are 
given by the fields $\psi^{[(i,1)(a_1,a_2)]}_{(a_2,a_2)}$ with $i=a_1+a_2+1-2l$, $1\le l\le N$. Here and elsewhere in the paper we put the boundary condition labels 
as superscripts in square brackets. 

The rest of the paper is organized as follows. After some preliminaries in section 2 we give a detailed formula for the RG pairing in section 3. 
In section 4 we analyse the flow $(2,2)\to (3,1)\oplus (1,1)$ for which we calculate the leading order expansions (\ref{prescription2}) for fields of dimensions 
near zero and near 1, that is for weights $h= {\cal O}(1/m)$ and $h= 1 + {\cal O}(1/m)$ respectively. The result is shown to match the expansions found by other methods in \cite{GRW}.
In section 5  we calculate the expansions of fields of dimension near 0 
for more general flows: $(p,2)\to (p-1,1)\oplus (p+1,1)$. In section 6 we check the first subleading corrections for the expansions found in section 5 against 
 conformal perturbation theory. We conclude with some brief comments in section 7. The appendices contain the expressions for OPE coefficients and 
 a derivation of asymptotic expansions for conformal blocks.

\section{Some preliminaries}
\setcounter{equation}{0}

The unitary Virasoro minimal models $M_{m}$ have central charges 
\be
c_{m}=1 - \frac{6}{m(m+1)} \, 
\ee
where $m\ge 3$ is an integer. We assume the diagonal modular invariant. The conformal weights of primaries 
are given by the values 
\be
h_{(r,s)}=\frac{((m+1)r-ms)^2-1}{4m(m+1)} 
\ee 
where $(r,s)$ belong to the Kac table: 
$(r,s)\in K=\{(r',s'): 1\le r'\le m-1, 1\le s'\le m\}$ defined modulo the symmetry $(r,s)\to (m-r,m+1-s)$.
For large values of $m$ we have 
\be
h_{(r,s)} = \frac{(r-s)^2}{4} + \frac{r^2-s^2}{4m} + \frac{s^2-1}{4m^2} + {\cal O}(m^{-3}) \, \, .
\ee
 
 The fusion rules for chiral fields $\phi_{(r,s)}$ are 
 \be
 \phi_{(r,s)}\times \phi_{(r',s')}= \sum_{r'',s''} {\cal N}_{(r,s)(r',s')}^{(r'',s'')}\phi_{(r'',s'')}\, , 
 \ee
 \be
 {\cal N}_{(r,s)(r',s')}^{(r'',s'')}= {\cal N}_{r,r'}^{r''}(m){\cal N}_{s,s'}^{s''}(m+1)\, , 
 \ee
 \be
 {\cal N}_{a,b}^{c}(m)=
 \left \{
\begin{array}{l@{\qquad}l}
1\, ,  &|a-b|+1 \le c \le  {\rm min}(a+b-1, 2m-a-b-1)\, , a+b+c \mbox{ odd } 
\\[1ex]
0\, , &\mbox{ otherwise} 
\end{array}
\right .
\ee
 For large values of $m$ we get  simpler $SU(2)$ fusion rules
 \be 
 \lim_{m\to \infty}{\cal N}_{a,b}^{c}(m)=
 \left \{
\begin{array}{l@{\qquad}l}
1\, ,  &|a-b|+1 \le c \le  a+b-1\, , a+b+c \mbox{ odd } 
\\[1ex]
0\, , &\mbox{ otherwise} 
\end{array}
\right .
 \ee

Irreducible conformal boundary conditions in $M_{m}$ are labeled by a pair from the Kac table: $(a_{1},a_{2})$. 
The spectrum of boundary fields $\psi^{[(a_{1},a_{2})(b_{1},b_{2})]}_{(r,s)}$ which join  two such boundary conditions 
is determined from the following decomposition of the state space 
\be
{\cal H}_{(a_{1},a_{2})(b_{1},b_{2})}= \bigoplus_{(c_{1},c_{2})\in K}{\cal N}_{(a_1,a_2)(b_1,b_2)}^{(c_1,c_2)}
{\cal H}_{(c_1,c_2)}\, . 
\ee

The OPE of fields $\psi^{[(a_{1},a_{2})(b_{1},b_{2})]}_{(i_1,i_2)}$ has the following form 
\be
 \psi^{[a b]}_{i}(x)\psi^{[b c]}_{j}(y) 
 \sim 
\sum_{l} C^{[a b c]\, l}_{ij}
\psi^{[a c]}_{l}(y)(x-y)^{h_{l}-h_{i}-h_{j}} \, , \enspace x>y
\ee
Here for brevity each index stands for a pair from the Kac table, e.g. $l=(l_1,l_2)\in K$.

The OPE coefficients for  (unnormalized) boundary fields can be expressed in terms of the fusion matrices \cite{Runkel}
\be \label{OPEunnormalized}
\tilde C^{[a b c]\, l}_{ij} =F_{bl}\left[\begin{array}{cc}a& c\\
i&j \end{array}  \right] 
\ee
These OPE coefficients satisfy the identities 
\be 
\tilde C_{ij}^{[abc]k}=\tilde C_{ji}^{[cba]k} \, , \quad
\tilde C_{ij}^{[abc]k}\tilde C_{kk}^{[aca]1}=\tilde C_{jk}^{[bca]i}\tilde C_{ii}^{[aba]1} \, . 
\ee 
Normalizing the fields $\psi^{[aa]}_{i}$ so that 
$$
\psi^{[aa]}_{i}(x) \psi^{[aa]}_{i}(y) = \frac{1}{(x-y)^{2h_{i}}}{\bf 1}_{aa} + \dots 
$$
we get the normalized OPE coefficients 
\be \label{OPE_norma}
C^{[a a a]\, k}_{ij} =F_{ak}\left[\begin{array}{cc}a& a\\
i&j \end{array}  \right]\left( \frac{F_{a1}\left[\begin{array}{cc}a& a\\
k&k \end{array}  \right]}{ F_{a1}\left[\begin{array}{cc}a& a\\
i&i \end{array}  \right] F_{a1}\left[\begin{array}{cc}a& a\\
j&j \end{array}  \right] }  \right)^{1/2} \, . 
\ee

More details on the OPE coefficients are given in appendix A.

\section{RG pairing}
\setcounter{equation}{0}
In this section we bring the general prescription (\ref{prescription1}), (\ref{prescription2}) 
to a more concrete form. We have in mind applications to Virasoro minimal models but most 
of the formulae below can be easily generalized to include other theories.   

To apply  (\ref{prescription1}), (\ref{prescription2}) we 
normalize the UV fields $\psi_{i}^{\rm UV}$ so that 
\be \label{UVnorm}
C_{ii}^{[{\rm UV\,  UV\,  UV}]1}= 1\, , \quad \langle \psi_{i}^{\rm UV}(x)\psi_{i}^{\rm UV}(y)\rangle 
= \frac{g_{\rm UV}}{(x-y)^{2h_{i}}}
\ee
where $g_{\rm UV}=\langle 1_{\rm UV} \rangle$ is the $g$-factor (ground state degeneracy, \cite{AL}) of the UV boundary condition.
(Here for simplicity we assume that the UV boundary condition is irreducible. )
Let us further consider the expansions
\be \label{RGdef}
\psi^{\rm IR, UV}(x) = \sum_{a} \xi_{a} \hat \psi_{a}(x) \, , \qquad 
\psi^{\rm UV, IR}(x) = \sum_{a} \xi_{a} \hat \psi_{ a}^{\dagger}(x) 
\ee 
where the index $a$ labels the irreducible components of the IR boundary condition and 
the fields
 $$\hat \psi_{a}\equiv \hat \psi^{[a, {\rm UV}]}_{a}\, , \quad  
\hat \psi_{a}^{\dagger}\equiv \hat \psi^{[ {\rm UV}, a]}_{a}$$ 
 are primaries in the corresponding 
boundary condition changing sectors which are normalized so that 
\be \label{hat_norm}
\langle \hat \psi_{a}(x)\hat \psi^{\dagger}_{a}(y)\rangle =  
\langle \hat \psi^{ \dagger}_{a}(x) \hat \psi_{a}(y) \rangle = \frac{1}{(x-y)^{2h_{a}}}\, . 
\ee

Using these  definitions we can write the decomposition into  conformal blocks 
for the RG-pairing (\ref{prescription1}) 
\bea \label{detailed_pairing}
\langle \psi_{j}^{[a,b]}, \psi_{i}^{\rm UV}\rangle_{\rm RG} &=& 
\langle \hat \psi_{a}^{\dagger}(\infty) \psi_{j}^{[a,b]}(1)\hat \psi_{b}\left(\frac{1}{2}\right)\psi_{i}^{\rm UV}(0)
\rangle \nonumber \\
&=&
 \xi_{a}\xi_{b} \sum_{p} C_{jp}^{[ab\, {\rm UV}]\hat \psi_{a}}
C_{\hat \psi_{b} i}^{[b\, {\rm UV}\, {\rm UV}]p}{\cal F}_{\hat a j ;\hat b i}^{p}\left(\frac{1}{2} \right)
\eea 
where the indices $\hat a$, $\hat b$ label the Virasoro representations corresponding to the fields 
$\hat \psi_{a}$.

For the flows 
\be 
(a_1,a_2)\longrightarrow \bigoplus_{i=1}^{N}(a_1+a_2 +1 -2i,1)\, , \quad N={\rm min}(a_1,a_2)
\ee
we label the components by the index $a\in \{a_{1}+a_{2}+1-2i|i=1,2,\dots N \}$. We propose that 
\be \label{pmin}
\hat \psi_{a} = \psi_{(a_2,a_2)}^{[(a,1)(a_1,a_2)]} \, 
\ee
with the normalization (\ref{hat_norm}). Thus we have $\hat a = \hat b = (a_{2},a_{2})$ in 
(\ref{detailed_pairing}) which can be now written as 
\be\label{detailed2}
\langle \psi_{j}^{[(a,1),(b,1)]}, \psi_{i}^{[{\rm UV}]}\rangle_{\rm RG} = \xi_a \xi_b 
\sum_{p} C_{jp}^{[(a,1)(b,1)(a_1,a_2)](a_2,a_2)}
C_{(a_2,a_2)i}^{[(b,1)(a_1,a_2)(a_1,a_2)]p}{\cal F}_{(a,1) j ;(b,1) i}^{p}\left(\frac{1}{2} \right)
\ee
where $i=(i_1,i_2)$, $j=(j_1,j_2)$, $p=(p_1,p_2)$. 

The pairing (\ref{detailed2}) is now expressed in terms of the OPE coefficients which can be calculated 
using fusion matrices (\ref{OPEunnormalized}) (see appendix A ), the minimal model conformal blocks ${\cal F}_{ij,kl}^{p}$ and the expansion coefficients $\xi_{a}$.
We will see in the forthcoming sections how one can fix the coefficients $\xi_{a}$ for particular examples of these flows.
 
 Our proposal (\ref{pmin}) is essentially a guess. We expect the RG defect to be close to the identity operator 
 so its dimension should go to zero as $m\to \infty$. Equation (\ref{pmin}) is the simplest possibility. 
 In addition we offer the following argument that further limits the choices. It is claimed in \cite{GrahamWatts} that 
 given a paricular boundary flow in minimal models one can apply to it the bulk topological 
 defect corresponding to the $(r,1)$ representation and get another boundary flow. It seems reasonable to 
 us that the statement of \cite{GrahamWatts} should also extend to the boundary defect operators. Namely, if one knows 
 the RG boundary defect field corresponding to the original flow the defect field for the image flow can be obtained 
 via the action of the same bulk topological defect. Consider then the  boundary flow triggered by $\psi_{13}$: 
 $
 (1,a_2) \to (a_2,1) \, 
 $.
 In this case there is only one boundary condition changing primary field between the UV and IR fixed points: 
 $\psi^{[(a_2, 1)(1,a_2)]}_{(a_2,a_2)}$. 
 Thus the RG boundary defect field must be built using this field and possibly its descendants.
 Since the fusion with a topological defect $(a_1,1)$ cannot change the representation content 
 we obtain that the RG boundary defect field for a flow from the $ (a_1,a_2)$ boundary condition must be 
 built upon the Virasoro representation $(a_2,a_2)$. There is only one such primary in each boundary sector 
 - the one given by formula (\ref{pmin}). The simplest possibility is then that the defect fields 
 are given by the appropriately normalized primary components.


\section{The flow $(2,2)\to (3,1) \oplus (1,1)$}
\setcounter{equation}{0}

In this section we focus on the most simple example of boundary flows considered in \cite{RRS} - 
the flow from the $(2,2)$ boundary condition into the superposition of $(3,1)$ and $(1,1)$ boundary 
conditions. We will investigate in detail the mapping of fields of dimensions near 0 and near 1. 
For $m=\infty$ a mapping of these fields was worked out in \cite{GRW}. We will reproduce   
their answers using our RG pairing (\ref{detailed_pairing}). We will also obtain a prediction for 
the finite values of $m$.

Let us now list the fields involved and fix the rest of normalizations. 
For the UV boundary condition the complete list of primaries is
\be
{\bf 1}_{22} \equiv \psi_{(1,1)}^{[(2,2)(2,2)]}\, , \enspace \phi\equiv \psi_{(3,3)}^{[(2,2)(2,2)]}\, , 
\enspace \psi =\psi_{(1,3)}^{[(2,2)(2,2)]}\, , \enspace \bar \psi=\psi_{(3,1)}^{[(2,2)(2,2)]}
\ee
where we use essentially the same notations as in \cite{GRW}. These fields are normalized as in (\ref{UVnorm}). 
Together with the  primaries $\psi$ and $\hat \psi$ there is also a descendant $\partial \phi$ which has 
a dimension near 1.  
As explained in \cite{GRW} to account for an apparent jump in the number of null vectors in the $m\to \infty$ 
limit one introduces a rescaled field 
\be\label{d3}
d_{3}(x) = - \frac{m}{2}\partial \phi(x) \, . 
\ee
Although the state $L_{-1}|(3,3)\rangle$ becomes null in the $m\to \infty$ limit the rescaled field $d_{3}$ 
retains a finite norm throughout and does not decouple. 

In the IR we have boundry fields 
\be
{\bf 1}_{11}\equiv\psi_{(1,1)}^{[(1,1)(1,1)]}\, , \enspace {\bf 1}_{31}\equiv\psi_{(1,1)}^{[(3,1)(3,1)]}\, , \enspace \varphi_{31}\equiv \psi_{(3,1)}^{[(3,1)(3,1)]}\, , \enspace 
\varphi_{51}\equiv \psi_{(5,1)}^{[(3,1)(3,1)]} \, , 
\ee
\be
\tilde \varphi_{31}\equiv \psi_{(3,1)}^{[(1,1)(3,1)]}\, , \quad 
\tilde \varphi_{31}^{\dagger} \equiv \psi_{(3,1)}^{[(3,1)(1,1)]}\, . 
\ee
The fields $1_{11},1_{31}, \varphi_{31}, \varphi_{51}$ are normalized similarly to (\ref{UVnorm}) while the fields 
$\tilde \varphi_{31}$, $\tilde \varphi_{31}^{\dagger}$ are normalized so that 
\be
\tilde \varphi_{31}(x) \tilde \varphi_{31}^{\dagger}(y) \sim \frac{1}{(x-y)^{2h_{31}}}{\bf 1}_{11} + \dots 
\ee

 We have the following fields of the type $\psi^{{\rm UV}, {\rm IR}},\psi^{{\rm IR}, {\rm UV}} $: 
 \be 
 \hat \psi_{1} \equiv  \psi^{[(1,1)(2,2)]}_{(2,2)}\, , \enspace  \hat \psi_{3} \equiv  \psi^{[(3,1)(2,2)]}_{(2,2)}\, 
 ,\enspace \hat \psi_{42}\equiv \psi^{[(3,1)(2,2)]}_{(4,2)}\, , 
 \ee
 \be 
 \hat \psi_{1}^{\dagger} \equiv  \psi^{[(2,2)(1,1)]}_{(2,2)}\, ,\enspace  \hat \psi_{3}^{\dagger} \equiv  
 \psi^{[(2,2)(3,1)]}_{(2,2)}\, 
 ,\enspace \hat \psi_{42}^{\dagger}\equiv \psi^{[(2,2)(3,1)]}_{(4,2)}\, . 
 \ee
These fields are normalized as in (\ref{hat_norm}). The fields $\hat \psi_{1}$ and $\hat \psi_{3}$ are 
the components of the RG defect field (\ref{RGdef}) while the field $\hat \psi_{42}$ arises as an 
intermediate channel in the conformal block decomposition (\ref{detailed_pairing}).

Since we are dealing with a perturbative RG flow we expect that only operators of nearby 
scaling dimensions get mixed. Thus we can analyze groups of operators with close conformal weights. 
We will only look at two groups: dimension near 0 and dimension near 1 operators. 
 
Starting with the operators of dimension near zero we  note  that the following RG pairings 
do not require knowledge of any nontrivial 
conformal blocks and can be expressed as  
\bea \label{easy_coefs}
\langle {\bf 1}_{11}, {\bf 1}_{22} \rangle_{\rm RG} &=& (\xi_{1}(m))^{2} \, , \quad 
 \langle {\bf 1}_{11}, \phi \rangle_{\rm RG} =  (\xi_{1}(m))^{2} \alpha_{1}
  \, , \nonumber \\
 \langle {\bf 1}_{31}, {\bf 1}_{22} \rangle_{\rm RG} &=& (\xi_{3}(m))^{2} \, , \quad 
 \langle {\bf 1}_{31}, \phi \rangle_{\rm RG} =  (\xi_{3}(m))^{2} \alpha_{3}  \, .
\eea
where 
\be 
\alpha_{1}=C^{[11,22,22]22}_{22,33}\, , \quad
\alpha_{3}=C^{[31,22,22]22}_{22,33} \, .
\ee
In these expressions we dropped the brackets  around the pairs of numbers labeling 
Virasoro representations as well as some commas. Thus $31$ stands for the $(3,1)$ representation. 
To avoid clutter we will use this shorthand notation  below in the OPE coefficients and conformal 
blocks whenever each number in a pair is a digit.  

The above expressions  hold for a finite $m$.
Here $\xi_{1}(m)$ and $\xi_{3}(m)$ are the coefficients in expansion (\ref{RGdef}) which depend on $m$. 
  To find the exact expressions for these coefficients we note that formulae (\ref{prescription2}), 
  (\ref{easy_coefs}) imply
  \be \label{projs}
  {\bf 1}_{11}= (\xi_{1}(m))^{2}( {\bf 1}_{22} + \alpha_{1}\phi) \, , \quad 
  {\bf 1}_{31}= (\xi_{3}(m))^{2}( {\bf 1}_{22} + \alpha_{3}\phi)\, . 
  \ee
The OPE algebra for ${\bf 1}_{11}$, ${\bf 1}_{31}$ is that of the projector operators: 
\be \label{1}
{\bf 1}_{11} \cdot {\bf 1}_{11} = {\bf 1}_{11}\, , \quad 
{\bf 1}_{31} \cdot {\bf 1}_{31} = {\bf 1}_{31}\, , \quad 
{\bf 1}_{31}\cdot {\bf 1}_{11} =0\, . 
\ee
On the other hand the relevant part of the deformed OPE algebra for the fields ${\bf 1}_{22}$, 
$\phi$ has the form 
\be \label{2}
\phi(x) \phi(0) \sim \frac{1}{x^{2\Delta_{\phi}(\lambda)}}{\cal D}_{\phi,\phi}^{1}(\lambda){\bf 1}_{22} + 
 \frac{1}{x^{\Delta_{\phi}(\lambda)}}{\cal D}_{\phi,\phi}^{\phi}(\lambda)\phi(0) 
+ \dots  
\ee 
where $\lambda$ is the coupling constant in front of the $\int \psi(x) dx$ perturbation, 
$\Delta_{\phi}(\lambda)$ is the deformed scaling dimension of $\phi$ and ${\cal D}_{\phi,\phi}^{1}(\lambda)$, 
${\cal D}_{\phi,\phi}^{\phi}(\lambda)$ are the deformed OPE coefficients. 
Let $\lambda^{*}$ be the value of the coupling at the IR fixed point. 
One can check perturbatively that $\Delta_{\phi}(\lambda^{*})=0$\footnote{The leading order perturbative calculation of the shift in 
anomalius dimension is subtle because one has to 
use a correction coming from a four point function. The details will be published elswhere \cite{withCor}.}.
Denoting ${\cal D}_{\phi,\phi}^{1}={\cal D}_{\phi,\phi}^{1}(\lambda^{*})$, ${\cal D}_{\phi,\phi}^{\phi}={\cal D}_{\phi,\phi}^{\phi}(\lambda^{*})$ we obtain from (\ref{easy_coefs}), (\ref{1}), (\ref{2}) 
\be 
{\cal D}_{\phi,\phi}^{1}=-\frac{1}{\alpha_{1}\alpha_{3}}\, , \qquad 
{\cal D}_{\phi,\phi}^{\phi}=-\Bigl(\frac{1}{\alpha_{1}} + \frac{1}{\alpha_{3}}\Bigr)\, , 
\ee
 \be\label{xis}
 (\xi_{1}(m))^{2}= \frac{1}{1-\frac{\alpha_{1}}{\alpha_{3}}}\, , \qquad 
 (\xi_{3}(m))^{2}= \frac{1}{1-\frac{\alpha_{3}}{\alpha_{1}}}\, . 
 \ee
The exact expressions for the coefficients $\alpha_{1}$, $\alpha_{3}$ in terms of Euler's Gamma functions 
are given in (\ref{alpha1exact}), 
(\ref{alpha3exact}). The expression for their ratio is particularly simple 
\be 
\frac{\alpha_{3}}{\alpha_{1}} = - \frac{\sin\left(\frac{\pi}{m}\right)}{\sin\left(\frac{3\pi}{m}\right)}\, . 
\ee
We notice that this coincides up to the sign with the ratio of boundary entropies of the two IR components
\be
\frac{\alpha_{3}}{\alpha_{1}} = -\frac{g_{11}}{g_{31}}
\ee
\bea
g_{11}&=& \left( \frac{8}{m(m+1)}\right)^{1/4} \left(\sin\left(\frac{\pi}{m}\right)\sin\left(\frac{\pi}{m+1}\right)      \right)^{1/2}\, , \nonumber \\
g_{31}&=&\left( \frac{8}{m(m+1)}\right)^{1/4} \frac{\sin\left(\frac{3\pi}{m}\right)\sin\left(\frac{\pi}{m+1}\right)}
{\left(\sin\left(\frac{\pi}{m}\right)\sin\left(\frac{\pi}{m+1}\right)\right)^{1/2}}\, . 
\eea
Thus the coefficients $\xi_{1}$, $\xi_{3}$ can be expressed in terms of the boundary entropies as 
\be 
(\xi_{1}(m))^{2}=\frac{g_{11}}{g_{11}+g_{31}}\, , \qquad 
(\xi_{3}(m))^{2}=\frac{g_{31}}{g_{11}+g_{31}}\, . 
\ee
We will show in section \ref{secp} that similar expressions hold for more general flows into two infrared components.
It is tempting to conjecture that similar expressions will hold for all flows of the type considered in 
\cite{RRS}. We postpone other checks of  this hypothesis to future work.

Asymptotically one has 
\bea 
\alpha_{1} &= &\sqrt{3}-\left(\frac{2\pi^2}{\sqrt{3}}\right)\frac{1}{m^2}+ \left(\frac{2\pi^2}{\sqrt{3}}\right)\frac{1}{m^3} 
+ {\cal O}(m^{-4}) \, , \nonumber \\
 \alpha_{3} &=&  -\frac{1}{\sqrt{3}} -\left(\frac{2\sqrt{3}\pi^2}{9}\right)\frac{1}{m^2} 
 -\left(\frac{2\sqrt{3}\pi^2}{9}\right)\frac{1}{m^3} + {\cal O}(m^{-4})
\eea
Substituting these expressions into (\ref{xis}), (\ref{projs}) we obtain at $m=\infty$ 
\be 
{\bf 1}_{11}= \frac{1}{4}( {\bf 1}_{22} + \sqrt{3}\phi) \, , \quad 
  {\bf 1}_{31}= \frac{3}{4}( {\bf 1}_{22} - \frac{1}{\sqrt{3}}\phi)\, , \quad m=\infty 
\ee
that matches with formula (3.31) in \cite{GRW}.


We next  calculate the  RG pairings  that involve the dimension near 1 fields: $ \varphi_{31}$, $\tilde \varphi_{31}$, $\tilde \varphi_{31}^{\dagger}$.
 These pairings involve contributions from 
conformal blocks and we will only work out the answer at  $m=\infty$. 

We have the following expressions for the RG pairings 
\be 
\langle \varphi_{31},\psi\rangle_{\rm RG} = (\xi_{3})^{2} C_{31,22}^{[31,31,22]22} 
C_{22,13}^{[31,22,22]22}
{\cal F}_{22,31;22,13}^{22}\left( \frac{1}{2}\right) \, , 
\ee
\bea
\langle \varphi_{31},\bar \psi\rangle_{\rm RG} &= &(\xi_{3})^{2} \Bigl[C_{31,22}^{[31,31,22]22} 
C_{22,31}^{[31,22,22]22}
{\cal F}_{22,31;22,31}^{22}\left( \frac{1}{2}\right) \nonumber \\
&& + C_{31,42}^{[31,31,22]22} C_{22,31}^{[31,22,22]42}
{\cal F}_{22,31;22,31}^{42}\left( \frac{1}{2}\right)\Bigr]\, , 
\eea
\bea
\langle \varphi_{31},d_3 \rangle_{\rm RG}  &= &-\frac{m}{2}(\xi_{3})^{2} \Bigl[C_{31,22}^{[31,31,22]22} C_{22,33}^{[31,22,22]22}
\tilde {\cal F}_{22,31;22,33}^{22}\left( \frac{1}{2}\right) \nonumber \\
&& + C_{31,42}^{[31,31,22]22} C_{22,33}^{[31,22,22]42}
\tilde {\cal F}_{22,31;22,33}^{42}\left( \frac{1}{2}\right)\Bigr]
\eea
where 
$$
\tilde {\cal F}_{ij;kl}^p$$ stand for conformal blocks in which  $L_{-1}\psi_{l}$ is inserted at the origin. 
We further have 
 \be 
 \langle \tilde \varphi_{31}, \psi \rangle_{\rm RG} = \xi_{1}\xi_{3}C_{31,22}^{[11,31,22]22}
 C_{22,13}^{[31,22,22]22}
 {\cal F}_{22,31;22,13}^{22}\left(\frac{1}{2}\right) \, , 
 \ee
 \bea
 \langle \tilde \varphi_{31}, \bar \psi \rangle_{\rm RG} = &= &\xi_{1}\xi_{3} \Bigl[C_{31,22}^{[11,31,22]22} C_{22,31}^{[31,22,22]22}
{\cal F}_{22,31;22,31}^{22}\left( \frac{1}{2}\right) \nonumber \\
&& + C_{31,42}^{[11,31,22]22} C_{22,31}^{[31,22,22]42}
{\cal F}_{22,31;22,31}^{42}\left( \frac{1}{2}\right)\Bigr]\, , 
 \eea
 \bea
\langle \tilde \varphi_{31},d_3 \rangle_{\rm RG}  &= &-\frac{m}{2}\xi_{1}\xi_{3} \Bigl[C_{31,22}^{[11,31,22]22} C_{22,33}^{[31,22,22]22}
\tilde {\cal F}_{22,31;22,33}^{22}\left( \frac{1}{2}\right) \nonumber \\
&& + C_{31,42}^{[11,31,22]22} C_{22,33}^{[31,22,22]42}
\tilde {\cal F}_{22,31;22,33}^{42}\left( \frac{1}{2}\right)\Bigr]\, , 
\eea
\be 
 \langle \tilde \varphi_{31}^{\dagger}, \psi \rangle_{\rm RG} 
 = \xi_{1}\xi_{3}C_{31,22}^{[31,11,22]22}C_{22,13}^{[11,22,22]22}{\cal F}_{22,31;22,13}^{22}
 \left(\frac{1}{2}\right) \, , 
 \ee
 \be 
 \langle \tilde \varphi_{31}^{\dagger}, \bar \psi \rangle_{\rm RG} 
 = \xi_{1}\xi_{3}C_{31,22}^{[31,11,22]22}C_{22,31}^{[11,22,22]22}{\cal F}_{22,31;22,31}^{22}
 \left(\frac{1}{2}\right) \, , 
 \ee
 \be 
 \langle \tilde \varphi_{31}^{\dagger}, d_3 \rangle_{\rm RG} 
 = -\frac{m}{2}\xi_{1}\xi_{3}C_{31,22}^{[31,11,22]22}C_{22,33}^{[11,22,22]22}
 \tilde {\cal F}_{22,31;22,33}^{22}
 \left(\frac{1}{2}\right)\, . 
 \ee
As shown in appendix B.1 the leading contributions from the conformal blocks are
\be \label{cblocks1}
{\cal F}_{22,31;22,13}^{22}\left( \frac{1}{2}\right) \sim -\frac{2}{3}m^{2} \, , \quad
{\cal F}_{22,31;22,31}^{22}\left( \frac{1}{2}\right) \sim  -\frac{2}{3}m^{2}\, , 
\ee
\be\label{cblocks2}
{\cal F}_{22,31;22,31}^{42}\left( \frac{1}{2}\right) \sim  1 \, , \quad 
\tilde {\cal F}_{22,31;22,33}^{22}\left( \frac{1}{2}\right)\sim \frac{4}{3}\, , \quad  \tilde {\cal F}_{22,31;22,33}^{42}\left( \frac{1}{2}\right) \sim -1 \, . 
 \ee
The leading order asymptotics for the OPE coefficients are given in formulae (\ref{firstOPE})-(\ref{lastOPE}). 
Using those formulae along with (\ref{cblocks1}), (\ref{cblocks2}) we obtain the leading order expansions
\be \label{res1}
 \varphi_{3} = \sqrt{\frac{3}{8}}(\bar \psi - \psi)\, , 
 \ee
 \be \label{res2}
 \tilde \varphi_{31}= -\frac{s}{4}[\psi + \bar \psi - \sqrt{2}d_{3}]\, , \quad 
 \tilde \varphi_{31}^{\dagger}= -\frac{s}{4}[\psi + \bar \psi + \sqrt{2}d_{3}]\, , \enspace m=\infty
 \ee
  where $s=-{\rm sign}(\xi_{1}\xi_{3}) $. These expansions  match with formula (3.34) from \cite{GRW}.
  More precisely the expansions given in (3.34) of \cite{GRW} are fixed up to two unknown constants 
  denoted by the authors as $\lambda_{2}$ and $\lambda_{3}$. These constants must satisfy 
  the  relation $\lambda_{2}\lambda_{3}= \frac{1}{16}$ derived from the OPE algebra.  
  Our result (\ref{res1}),(\ref{res2}) corresponds to  the values $\lambda_{2}=\lambda_{3}=-\frac{s}{4}$.

\section{Flows from $(p,2)$ boundary conditions}\label{secp}
\setcounter{equation}{0}
In this section we will consider the RG flows
\be
(p,2)\longrightarrow (p-1,1)\oplus (p+1,1)
\ee
with $p>2$. We will focus on the dimension near zero sector. The results obtained will be further used 
in the analysis of $1/m$ corrections. The normalized UV fields of dimension near zero are 
${\bf 1}_{p,2}$ and $\phi=\psi_{(3,3)}^{[(p,2)(p,2)]}$. In the IR we have ${\bf 1}_{p-1,1}$,  ${\bf 1}_{p+1,1}$.
The normalized RG defect fields are 
$$\hat \psi_{p-1} = \psi_{(2,2)}^{[(p-1,1)(p,2)]}\, , \quad  
\hat \psi_{p+1} = \psi_{(2,2)}^{[(p+1,1)(p,2)]}\, . 
$$
Similarly to (\ref{easy_coefs}) we now have 
\bea \label{easy_coefs2}
\langle {\bf 1}_{p-1,1}, {\bf 1}_{p,2} \rangle_{\rm RG} &=& (\xi_{p-1}(m))^{2} \, , \quad 
 \langle {\bf 1}_{p-1,1}, \phi \rangle_{\rm RG} =  (\xi_{p-1}(m))^{2} \alpha_{p-1}
  \, , \nonumber \\
 \langle {\bf 1}_{p+1,1}, {\bf 1}_{p,2} \rangle_{\rm RG} &=& (\xi_{p+1}(m))^{2} \, , \quad 
 \langle {\bf 1}_{p+1,1}, \phi \rangle_{\rm RG} =  (\xi_{p+1}(m))^{2} \alpha_{p+1}  \, 
\eea
where 
\be 
\alpha_{p-1}=C^{[(p-1,1)(p,2)(p,2)]22}_{22,33}\, , \quad
\alpha_{p+1}=C^{[(p+1,1)(p,2)(p,2)]22}_{22,33} \, .
\ee
The asymptotic expansions for these OPE coefficients are given in (\ref{A1}), (\ref{A2}).

Following the same steps as in the previous section we obtain 
 \be \label{tocheck1}
{\cal D}_{\phi,\phi}^{1}=-\frac{1}{\alpha_{p-1}\alpha_{p+1}}\, , \qquad 
{\cal D}_{\phi,\phi}^{\phi}=-\Bigl(\frac{1}{\alpha_{p-1}} + \frac{1}{\alpha_{p+1}}\Bigr)\, , 
\ee
 \be\label{xisp}
 (\xi_{p-1}(m))^{2}= \frac{1}{1-\frac{\alpha_{p-1}}{\alpha_{p+1}}}\, , \qquad 
 (\xi_{p+1}(m))^{2}= \frac{1}{1-\frac{\alpha_{p+1}}{\alpha_{p-1}}}\, . 
 \ee
 Using (\ref{ratio_p}) and (\ref{gp}) we have
 \be\label{xisp2}
 (\xi_{p-1}(m))^{2}= \frac{g_{p-1}}{g_{p-1} + g_{p+1}}\, , \qquad 
 (\xi_{p+1}(m))^{2}= \frac{g_{p+1}}{g_{p-1} + g_{p+1}}\, 
 \ee
 where $g_{p\pm 1}$ are the boundary entropies of the IR components. 
 We have the following asymptotics 
 \be \label{alpha_asympt}
 \alpha_{p-1} =  \sqrt{\frac{p+1}{p-1}}\Bigl( 1 - \frac{\pi^2 p}{3m^2} + {\cal O}(m^{-3})\Bigr) \, , \quad 
 \alpha_{p+1}=-\sqrt{\frac{p-1}{p+1}}\Bigl( 1 + \frac{p\pi^2 }{3m^2} + {\cal O}(m^{-3})\Bigr) \, . 
 \ee
 Hence at $m=\infty$ we obtain the following expansions
 \be
 {\bf 1}_{p-1,1} = \frac{p-1}{2p}\Bigl[ {\bf 1}_{p,2} + \sqrt{\frac{p+1}{p-1}}\phi \Bigr]\, , \quad 
 {\bf 1}_{p+1,1} = \frac{p+1}{2p}\Bigl[ {\bf 1}_{p,2} - \sqrt{\frac{p-1}{p+1}}\phi \Bigr]
 \ee
 that matches with formulae (A.27), (A.28) of \cite{GRW}.

\section{Leading $1/m$ corrections}
\setcounter{equation}{0}
We are interested in checking some of the $1/m$ subleading terms against the RG calculations. 
We do this for equations (\ref{tocheck1}). Recall that the coefficients ${\cal D}_{\phi,\phi}^{1}(\lambda)$
and ${\cal D}_{\phi,\phi}^{\phi}(\lambda)$ are defined to be the OPE coefficients in the theory 
deformed by the perturbation $\lambda \int dx \psi(x)$. Renormalization subtracts the 
logarithmic divergences arising from  short distances in the $m\to \infty$ limit. For finite $m$ they 
manifest itselves as poles in anomalous dimensions, or, equivalently, terms divergent in $m$. 
The one-loop beta function is 
\be
\beta(\lambda) = (1-h_{13})\lambda + D\lambda^2\, , \quad D=C_{13,13}^{[(2,p)(2,p)(2,p)]13} \, .
\ee
The fixed point is at 
\be\label{lstar}
\lambda^{*}=\frac{h_{13}-1}{D} = \frac{8}{\sqrt{6}m} + {\cal O}\left(\frac{1}{m^2}\right)\, . 
\ee
The terms in the beta function from two loops and higher result in subleading $1/m$ corrections 
so that the leading asymptotics of $\lambda^{*}$ is fixed by (\ref{lstar}).

Equations (\ref{tocheck1}) predict the following asymptotics for the OPE 
coefficients at the IR fixed point: 
\bea\label{tocheck2}
{\cal D}_{\phi,\phi}^{1}\equiv  {\cal D}_{\phi,\phi}^{1}(\lambda^{*})&=&1 + {\cal O}\left(\frac{1}{m^{3}} \right)\, , 
\nonumber \\ 
{\cal D}_{\phi,\phi}^{\phi}\equiv {\cal D}_{\phi,\phi}^{\phi}(\lambda^{*})&=&\frac{2}{\sqrt{p^2-1}} - \frac{2\pi^2p^2}{3\sqrt{p^2-1}m^2} + {\cal O}\left(\frac{1}{m^3} \right)\, .
\eea
The perturbation theory expansions have the form 
\bea \label{pert_exps}
{\cal D}_{\phi,\phi}^{1}(\lambda) &=& 1 + D_{\phi,\phi}^{1(1)}\lambda + D_{\phi,\phi}^{1(2)}\lambda^2 + \dots \, , 
\nonumber \\
{\cal D}_{\phi,\phi}^{\phi}(\lambda) &=& D_{33,33}^{[(p,2)(p,2)(p,2)]33} + D_{\phi,\phi}^{\phi(1)}\lambda + D_{\phi,\phi}^{\phi(2)}\lambda^2 + \dots
\eea
From (\ref{lstar}) we see that terms of the order $m^{-1}$ and $m^{-2}$ in the expansions (\ref{tocheck2}) 
can come only from the terms written out  in  (\ref{pert_exps}).
Comparing  (\ref{333333}) with (\ref{tocheck2}), (\ref{pert_exps}) we obtain the following predictions 
for perturbative corrections: 
\be
D_{\phi,\phi}^{1(1)} \lambda^{*} + D_{\phi,\phi}^{1(2)}\cdot(\lambda^{*})^2 = {\cal O}\left(\frac{1}{m^3} \right)\, , 
\ee
 \be
D_{\phi,\phi}^{\phi(1)} \lambda^{*} + D_{\phi,\phi}^{\phi(2)}\cdot (\lambda^{*})^2 = {\cal O}\left(\frac{1}{m^3} \right)\, . 
\ee

In the rest of this section we show that these identities hold. 
To begin with it is easy to argue that 
\be \label{argue}
D_{\phi,\phi}^{1(2)} = {\cal O}\left(\frac{1}{m}\right)\, , \qquad  
D_{\phi,\phi}^{\phi(2)} = {\cal O}\left(\frac{1}{m}\right) \, . 
\ee
Such second order corrections come from the integrals 
\be 
\iint\!\! dx_1 dx_2 \langle \phi(z) \phi(0) \psi(x_1)\psi(x_2) \rangle \, , \quad 
\iint\!\! dx_1 dx_2 \langle \phi(z) \phi(0) \phi(z') \psi(x_1)\psi(x_2) \rangle \, . 
\ee
On the $(p,2)$ boundary conditions the OPE of $\psi$ and $\phi$ contains only $\psi$ 
and the corresponding OPE coefficient (\ref{333313}) goes as $m^{-1}$. Thus  the correlation 
functions at hand contain a factor of  $m^{-1}$ as well. Short distance divergences are 
subtracted by renormalization and hence any possible $m\to \infty$ divergences are subtracted as well. 
We conclude that (\ref{argue}) holds. 

Thus we need to show 
\be
D_{\phi,\phi}^{1(1)}={\cal O}\left(\frac{1}{m^2}\right)\, , \qquad 
D_{\phi,\phi}^{\phi(1)}={\cal O}\left(\frac{1}{m^2}\right) \, . 
\ee 

We follow the method of calculating perturbative corrections to OPE coefficients presented in \cite{GuidaMagnoli}. 
It is based on the action principle according to which 
\be 
\frac{\partial}{\partial \lambda} \langle \Phi_{1}(t_1)\dots \Phi_{n}(t_{n}) \rangle_{\lambda} 
= \int\!\! dx \langle \psi(x) \Phi_{1}(t_1)\dots \Phi_{n}(t_{n})  \rangle_{\lambda} 
\ee
for any correlator of renormalized operators $\Phi_{i}$ in the deformed theory. 

Consider the deformed correlator $\langle \phi(x)\phi(0)\rangle_{\lambda}$ where $x>0$ and
$\phi$ stands for a renormalized operator. The operator product expansion has the form 
(\ref{2}). Using this OPE inside the two-point function at hand, taking a derivative 
with respect to $\lambda$ and setting $\lambda=0$ afterwards we obtain 
\be
\int\!\! dt \langle \phi(x)\phi(0) \psi(t)\rangle = \frac{D_{\phi,\phi}^{1(1)}}{x^{2h_{33}}} 
- 2\partial \Delta_{\phi}(0)\frac{\ln x}{x^{2h_{33}}} + \dots
\ee 
where the ellipsis stands for terms less singular in the $x\to 0$ limit. 
Taking the integral of the three point function we obtain 
\be
\int\!\! dt \langle \phi(x)\phi(0) \psi(t)\rangle = \frac{C_{13,33}^{[(p,2)(p,2)(p,2)]33}}{x^{2h_{33}+h_{13}-1}}
\frac{\sqrt{\pi}\Gamma\left(h_{13}-\frac{1}{2} \right)\Gamma^2\left(h_{13}-\frac{1}{2} \right)}
{\Gamma^2\left(\frac{h_{13}}{2} \right)\Gamma\left(1-h_{13} \right)} \, .
\ee
We have the following $1/m$ expansion 
\be \label{int_expa}
\frac{1}{x^{2h_{33}+h_{13}-1}}
\frac{\sqrt{\pi}\Gamma\left(h_{13}-\frac{1}{2} \right)\Gamma^2\left(h_{13}-\frac{1}{2} \right)}
{\Gamma^2\left(\frac{h_{13}}{2} \right)\Gamma\left(1-h_{13} \right)} = 2m + 4\ln(x) + {\cal O}\left(\frac{1}{m}\right)\, .
\ee
Renormalization amounts to subtracting the linear divergence in $m$. Using (\ref{333313}) we obtain 
\be\label{DDas}
D_{\phi,\phi}^{1(1)} = {\cal O}\left(\frac{1}{m^2} \right)  \, , \quad
 \partial \Delta_{\phi}(0) = -2C_{13,33}^{[(p,2)(p,2)(p,2)]33} + {\cal O}\left(\frac{1}{m^2}\right) \, .
 \ee
Note that we are  expanding in $1/m$ the OPE coeficient $C_{13,33}^{[(p,2)(p,2)(p,2)]33}$ only {\it after} we 
have subtracted the linear term in (\ref{int_expa}).

The same (leading order) result can be obtained by introducing a short distance cut-off into the three point function, 
taking the $m=\infty$ limit for the integrand first, taking the integral 
 and then subtracting the logarithmic divergences\footnote{Using finite $m$ as regularization we obtain power 
 divergences in $m$ as $m\to \infty$. Using a short distance cutoff $\epsilon$  we obtain power 
 divergences in $\epsilon$. Keeping $\epsilon$ finite and taking $m\to \infty$ produces logarithmic divergences. 
 Subtracting such logarithms is equivalent to subtracting power divergences in $m$ up to finite terms of order 
 $1/m$}.     The short distance cut off approach is 
 in general  computationally simpler.  
 
 To analyze the $D_{\phi,\phi}^{\phi(1)}$ correction we start with the  three point function of 
 deformed theory
 $$
 \langle \phi(x')\phi(x)\phi(0)\rangle_{\lambda} \, , \quad x>0\, , \enspace x'>0 \, . 
 $$ 
 Taking $x\to 0$ we can use the deformed OPE (\ref{2}). Differentiating with respect to $\lambda$ and setting 
 $\lambda=0$ we obtain 
 \bea \label{principal}
 \int\!\! dt \langle \phi(x')\phi(x)\phi(0)\psi(t)\rangle && = \frac{D_{\phi,\phi}^{\phi(1)}}{x^{h_{33}}(x')^{2h_{33}}}
 + \frac{C_{33,33}^{[(p,2)(p,2)(p,2)]33}}{x^{h_{33}}}\int\!\! dt \langle \phi(x') \phi(0) \psi(t)\rangle
 \nonumber \\
 &&-\ln x \, \partial \Delta_{\phi}(0)\frac{C_{33,33}^{[(p,2)(p,2)(p,2)]33}}{x^{h_{33}}(x')^{2h_{33}}} + \dots
 \eea
 where the ellipsis stands for terms less singular in $x$.

 Denote 
 \be
 G(x) = \langle \phi(\infty)\phi(1)\phi(x) \psi(0)\rangle \, , \qquad \eta = \frac{(x'-t)x}{(x'-x)t} \, .
 \ee
 Using global conformal transformations we obtain 
 \be\label{full_corr}
 \langle \phi(x')\phi(x)\phi(0)\psi(t)\rangle =  \left \{
\begin{array}{l@{\qquad}l}
f_{1}(x,x',t)
G\left(\frac{1}{1-\eta}\right)\, ,   &t<0 \mbox{ or } t>x' 
\\[1ex]
 f_{2}(x,x',t)  
 G\left(1-\frac{1}{\eta}\right)\, ,  & 0<t<x 
 \\[1ex]
 f_{3}(x,x',t)  
 G\left(\eta \right)\, ,  & x<t<x' 
\end{array}
\right .
 \ee 
 where 
 \bea
 f_{1}(x,x',t) &=& (x'-t)^{h_{33}-h_{13}}(x'-x)^{h_{13}-h_{33}}(x-t)^{-h_{33}-h_{13}}(x')^{-2h_{33}}\, , \\
 f_{2}(x,x',t)&=& t^{h_{33}-h_{13}}(x')^{h_{13}-h_{33}}(x'-t)^{-h_{13}-h_{33}}x^{-2h_{33}}\, , \\
 f_{3}(x,x',t)&=&(t-x)^{h_{33}-h_{13}}x^{h_{13}-h_{33}}t^{-h_{33}-h_{13}}(x'-x)^{-2h_{33}} \, .
 \eea
 
  The four-point function $G(x)$  decomposes into conformal blocks as
  \be
  G(x) = X_{33}{\cal F}_{33,33;33,13}^{33}(x)      + X_{31}{\cal F}_{33,33;33,13}^{31}(x)  
  \ee 
  where (see formulas (\ref{333333}) - (\ref{331331}))
  \bea \label{XX}
  X_{33}&=& C_{33,33}^{[(p,2)(p,2)(p,2)]33}C_{33,13}^{[(p,2)(p,2)(p,2)]33}= \frac{8}{\sqrt{6(p^2-1)} m} + 
  {\cal O}\left(\frac{1}{m^2} \right) \, , \nonumber \\
  X_{31}&=& C_{33,33}^{[(p,2)(p,2)(p,2)]31}C_{33,13}^{[(p,2)(p,2)(p,2)]31} = \frac{4}{\sqrt{6(p^2-1)} m} + 
  {\cal O}\left(\frac{1}{m^2} \right) \, .
  \eea
  
  It is shown in appendix B.2 that the leading asymptotics for the conformal blocks are
  \be\label{FF}
  {\cal F}_{33,33;33,13}^{33}(x) = \frac{1}{x}-\frac{1}{2} + {\cal O}(\frac{1}{m})\, , \quad 
  {\cal F}_{33,33;33,13}^{31}(x)  = 1 + {\cal O}(\frac{1}{m}) \, . 
  \ee
  Using (\ref{XX}), (\ref{FF}) we obtain the leading asymptotics 
  \be 
   G(x) = \left(\frac{1}{x}\right) \frac{8}{\sqrt{6(p^2-1)} m} +  {\cal O}\left(\frac{1}{m^3} \right) \, . 
  \ee
From this and (\ref{full_corr}) we obtain
 \be\label{full_corr2}
 \langle \phi(x')\phi(x)\phi(0)\psi(t)\rangle =  \left \{
\begin{array}{l@{\qquad}l}
X_{33}\frac{x'}{(x'-t)(-t)} + {\cal O}\left(\frac{1}{m^2}\right) \, ,   &t<0 \mbox{ or } t>x' 
\\[1ex]
 X_{33}\frac{x}{(x-t)(t)} + {\cal O}\left(\frac{1}{m^2}\right)  \, ,  & 0<t<x 
 \\[1ex]
 X_{33}\frac{(x'-x)}{(t-x)(x'-t)} + {\cal O}\left(\frac{1}{m^2}\right) 
\, ,  & x<t<x' \, \enspace .
\end{array}
\right .
 \ee 
 We now proceed by using the short distance cut off version of renormalization as that is 
 more computationally concise.
Substituting (\ref{full_corr2})  into the integral on the left hand side of (\ref{principal}) regulated by 
a short distance cut off we obtain 
\bea
&& \int\!\! dt \theta(|t|-\epsilon)\theta(|t-x|-\epsilon)\theta(|t-x'|-\epsilon)\langle \phi(x')\phi(x)\phi(0)\psi(t)\rangle 
\nonumber \\
&& = X_{33}\Bigl[-6\ln\epsilon +2\ln(x'+\epsilon) +2\ln(x-\epsilon) + 2\ln(x'-x-\epsilon)  \Bigr]  + {\cal O}\left(\frac{1}{m^2}\right)
\eea
Renormalization amounts to (minimally) subtracting the $\ln\epsilon$ divergences. We thus obtain  
\be \label{corr_xx}
\int\!\! dt \langle \phi(x')\phi(x)\phi(0)\psi(t)\rangle = 2X_{33}\Bigl[\ln(x') +\ln(x) + \ln(x'-x)  \Bigr]  + {\cal O}\left(\frac{1}{m^2}\right) \, .
\ee
We also have 
\bea
&&\int\!\! dt \theta(|t|-\epsilon)\theta(|t-x'|-\epsilon) \langle \phi(x') \phi(0) \psi(t)\rangle 
= \nonumber \\
&& C_{33,13}^{[(p,2)(p,2)(p,2)]33}\Bigl[2\ln(x'-\epsilon) + 2\ln(x'+\epsilon) -4\ln\epsilon  \Bigr] 
+ {\cal O}\left(\frac{1}{m^2}\right)
\eea
Renormalizing and using (\ref{DDas}) we obtain 
\bea
&&\frac{D_{\phi,\phi}^{\phi(1)}}{x^{h_{33}}(x')^{2h_{33}}}
 + \frac{C_{33,33}^{[(p,2)(p,2)(p,2)]33}}{x^{h_{33}}}\int\!\! dt \langle \phi(x') \phi(0) \psi(t)\rangle
 -\ln x \, \partial \Delta_{\phi}(0)\frac{C_{33,33}^{[(p,2)(p,2)(p,2)]33}}{x^{h_{33}}(x')^{2h_{33}}} \nonumber \\
 && = D_{\phi,\phi}^{\phi(1)} + 4X_{33}\ln(x') + 2X_{33}\ln(x) + {\cal O}\left(\frac{1}{m^2}\right)\, .
\eea
Matching this with the leading asymptotics of (\ref{corr_xx}) in the $x\to 0$ limit we finally obtain 
the desired result
\be
D_{\phi,\phi}^{\phi(1)} = {\cal O}\left(\frac{1}{m^2}\right)\, . 
\ee

\section{Conclusion}
Here we summarize the main results of the paper and spell out some open questions. Mimicing Gaiotto's construction \cite{Gaiotto} 
we have proposed a general formula for RG pairings for boundary flows (\ref{prescription1}). It is proposed that 
all information about the mapping of fields is encoded in a special local boundary field - boundary RG defect.   For boundary flows 
in minimal models triggered by the field $\psi_{13}$ we propose a candidate for such a field (\ref{RGdef}), (\ref{pmin}) which we fix up to relative normalisation 
coefficients $\xi_{a}$. For flows from $(p,2)$ boundary conditions   we fix these coefficients up to  signs (see (\ref{xisp2})). Formula (\ref{xisp2}) is suggestive of 
a general relation in which the squares of coefficients $\xi_{a}$ are given by the ratio of the $g$-factor of the $a$-th IR component to the total IR $g$-factor. This conjecture needs further checks which we hope to perform in the future. For flows from $(2,2)$ boundary condition we found using our RG pairing 
the mapping of fields of dimensions near 0 and 1. The results at the leading order coincide with those obtained in \cite{GRW} by other methods. Our prescription 
gives the expansion to all orders in $1/m$. The terms subleading in $1/m$ should capture the RG corrections to OPE coefficients. In section 6 we checked some 
particular coefficients in the expansions of dimension near 0 fields against conformal perturbation calculations  at the first subleading order. We found a precise match. 
It would be desired to perform more checks of this kind. For that one would need to develop some systematics 
for the $1/m$ expansion of the relevant conformal 
blocks. It is conceivable that there are corrections to our RG defect fields $\hat \psi_{a}$ proportional to fields 
of higher dimension. In view of our remarks in the end of section 3 we believe that such fields can only be descendants 
of $\hat \psi_{a}$.  Such
 corrections must be suppressed by powers of $1/m$ but may enter into the game at higher orders. 

In this paper we limited ourselves to boundary flows triggered by the $\psi_{13}$ field which start from a single Cardy boundary condition.
It would be interesting to find candidates for boundary RG defects for other known boundary flows. Work on this and other related questions is currently underway.

\setcounter{equation}{0}

\setcounter{equation}{0}

\begin{center}
{\bf Acknowledgments}
\end{center}

The author wants to thank Cornelius Schmidt-Colinet for discussions and help with calculations of fusion matrices. 
He also thanks Daniel Friedan for inspiring discussions. This work was supported by STFC grant 
ST/J000310/1 "High energy physics at the Tait Institute".


\appendix
\renewcommand{\theequation}{\Alph{section}.\arabic{equation}}
\setcounter{equation}{0}
\section{Fusion matrices and structure constants}
It was explained in \cite{Runkel} that that for the A-series Virasoro minimal models 
the boundary structure constants in a certain normalization are given by the fusion matrices 
\be \label{OPE_unn}
\tilde C^{[a b c]\, l}_{ij} =F_{bl}\left[\begin{array}{cc}a& c\\
i&j \end{array}  \right] \, . 
\ee
In this paper we work with fields normalized as in (\ref{UVnorm}), (\ref{hat_norm}). The 
corresponding OPE coefficients differ from (\ref{OPE_unn}) by a normalization factors 
which are also expressible via the fusion matrices as in formula (\ref{OPE_norma}).

The fusion matrices can be calculated recursively as explained in \cite{Runkel} and are expressed via 
Euler's Gamma functions. We are indebted to C.~Schmidt-Colinet for the use of his computer code implementing 
the recursive procedure of \cite{Runkel}. Below we list the asymptotic values and expansions of various 
OPE coefficients used to calculate the RG pairings.  

\subsection{Flows from $(2,2)$ boundary conditions}

We record the following normalization coefficients with their leading asymptotics in the $m\to \infty$ limit
\be 
d_{13}=\left(F_{(22)(11)}\left[\begin{array}{cc}(22)& (22)\\
(13)&(13) \end{array}  \right]\right)^{-1/2} \sim \sqrt{\frac{3}{2}}\frac{1}{m} \, , 
\ee
\be
d_{31}=\left(F_{(22)(11)}\left[\begin{array}{cc}(22)& (22)\\
(31)&(31) \end{array}  \right]\right)^{-1/2} \sim \sqrt{\frac{3}{2}}\frac{1}{m} \, , 
\ee
\bea
d_{33}&& =\left(F_{(22)(11)}\left[\begin{array}{cc}(22)& (22)\\
(33)&(33) \end{array}  \right]\right)^{-1/2} \nonumber \\
&& = \left(-\frac{1}{4}\frac{m(m+1)\Gamma\left(\frac{3}{m+1}\right)\Gamma^{2}\left(1+\frac{2}{m}\right)
\Gamma\left(1-\frac{3}{m}\right)\sin\left(\frac{2\pi}{m}\right) }{\Gamma^{2}\left(\frac{2}{m+1}\right) 
\Gamma\left(-\frac{1}{m+1}\right)\Gamma\left(1+\frac{1}{m}\right)\sin\left(\frac{2\pi}{m+1}\right) } 
\right)^{-1/2}\nonumber \\
&& =\sqrt{3}-\left(\frac{2\pi^2}{\sqrt{3}}\right)\frac{1}{m^2}+ \left(\frac{2\pi^2}{\sqrt{3}}\right)\frac{1}{m^3} 
+ {\cal O}(m^{-4})\, , 
\eea
\be
\tilde d_{31}=\left(F_{(31)(11)}\left[\begin{array}{cc}(31)& (31)\\
(31)&(31) \end{array}  \right]\right)^{-1/2} \sim 2 \, , 
\ee
\be 
\mu = \left(F_{(31)(11)}\left[\begin{array}{cc}(22)& (22)\\
(42)&(42) \end{array}  \right]\right)^{1/2}\cdot
\left(F_{(31)(11)}\left[\begin{array}{cc}(22)& (22)\\
(22)&(22) \end{array}  \right]\right)^{-1/2} \sim  \frac{\sqrt{3}}{2m} \, , 
\ee
\be 
\nu = \left(F_{(11)(11)}\left[\begin{array}{cc}(22)& (22)\\
(22)&(22) \end{array}  \right]\right)^{1/2}\cdot 
\left(F_{(31)(11)}\left[\begin{array}{cc}(22)& (22)\\
(22)&(22) \end{array}  \right]\right)^{-1/2} \sim  \frac{1}{\sqrt{2}m} \, . 
\ee
We next list the normalized OPE coefficients involving the UV fields and the $\hat \psi_{a}$ fields  
\be \label{firstOPE}
C_{22,13}^{[31,22,22]22} = d_{13}F_{(22)(22)}\left[\begin{array}{cc}(31)& (22)\\
(22)&(13) \end{array}  \right] \sim \sqrt{\frac{3}{2}}\cdot \frac{1}{m}\, , 
\ee
\be 
C_{22,31}^{[31,22,22]22} = d_{31}F_{(22)(22)}\left[\begin{array}{cc}(31)& (22)\\
(22)&(31) \end{array}  \right] \sim -\frac{1}{m\sqrt{6}}\, , 
\ee
\be 
C_{22,31}^{[31,22,22]42} = d_{31}\mu F_{(22)(42)}\left[\begin{array}{cc}(31)& (22)\\
(22)&(31) \end{array}  \right] \sim -\frac{2\sqrt{2}}{3}\, , 
\ee
\bea \label{alpha3exact}
C_{22,33}^{[31,22,22]22} &&= d_{33} F_{(22)(22)}\left[\begin{array}{cc}(31)& (22)\\
(22)&(33) \end{array}  \right] =-d_{33}\left(\frac{\sin\left(\frac{\pi}{m}\right)}
{\sin\left(\frac{3\pi}{m}\right)}\right)
\nonumber \\
&& \sim -\frac{1}{\sqrt{3}} -\left(\frac{2\sqrt{3}\pi^2}{9}\right)\frac{1}{m^2} 
 -\left(\frac{2\sqrt{3}\pi^2}{9}\right)\frac{1}{m^3} + {\cal O}(m^{-4})\, , 
\eea
\be 
C_{22,33}^{[31,22,22]42} = d_{33}\mu F_{(22)(42)}\left[\begin{array}{cc}(31)& (22)\\
(22)&(33) \end{array}  \right] \sim  \frac{4}{3m}\, , 
\ee
\be \label{alpha1exact}
C_{22,13}^{[11,22,22]22} = d_{13}\, , \quad C_{22,33}^{[11,22,22]22}=d_{33}\, , \quad 
C_{22,31}^{[11,22,22]22} = d_{31}\, . 
\ee

The vertices involving the IR fields and the RG defect fields are as follows 
\be
C_{31,22}^{[31,31,22]22}=\tilde d_{31}F_{(31)(22)}\left[\begin{array}{cc}(31)& (22)\\
(31)&(22) \end{array}  \right] \sim \frac{1}{m}\, , 
\ee
\be
C_{31,42}^{[31,31,22]22}=\tilde d_{31}\mu^{-1} F_{(31)(22)}\left[\begin{array}{cc}(31)& (22)\\
(31)&(42) \end{array}  \right] \sim -\frac{1}{\sqrt{3}}\, , 
\ee
\be
C_{31,22}^{[11,31,22]22}=\nu   \, , \quad C_{31,42}^{[11,31,22]22}=\nu  \mu^{-1}\, , 
\ee
\be \label{lastOPE}
C_{31,22}^{[31,11,22]22}=\nu^{-1} F_{(11)(22)}\left[\begin{array}{cc}(31)& (22)\\
(31)&(22) \end{array}  \right] \sim \frac{1}{\sqrt{2}m}\, . 
\ee

\subsection{Flows from $(p,2)$ boundary conditions}
We have the following normalization factor for the $\psi_{(3,3)}^{[(p,2)(p,2)]}$ field 
\bea 
d_{33}(p)&& = \left( F_{(p,2)(11)}\left[\begin{array}{cc}(33)& (33)\\
(p,2)&(p,2) \end{array}  \right]  \right)^{-1/2} \nonumber \\
&& \sim \sqrt{\frac{p+1}{p-1}}\Bigl( 1 -\frac{1}{3m^{2}}[-6\gamma + \pi^2 p 
-6(\psi(p-2)+(p-2)\psi'(p-2))] \Bigr)
\eea
where the expansion is up to the terms of order $m^{-3}$. For the $\psi_{(1,3)}^{[(p,2)(p,2)]}$ and 
$\psi_{(3,1)}^{[(p,2)(p,2)]}$ fields 
the similar factors are
\bea 
d_{13}(p)&& = \left( F_{(p,2)(11)}\left[\begin{array}{cc}(13)& (13)\\
(p,2)&(p,2) \end{array}  \right]  \right)^{-1/2} \nonumber \\
&& \sim \frac{p-2}{\sqrt{6}} \, , \enspace \mbox{ for } p>2 \, ,  
\eea
\bea 
d_{31}(p)&& = \left( F_{(p,2)(11)}\left[\begin{array}{cc}(31)& (31)\\
(p,2)&(p,2) \end{array}  \right]  \right)^{-1/2} \nonumber \\
&& \sim (p-2)\sqrt{\frac{p+1}{2(p-1)}} \, , \enspace \mbox{ for } p>2 \, .  
\eea

We record the following expansions for normalized OPE coefficients 
\be \label{A1}
C^{[(p-1,1)(p,2)(p,2)]22}_{22,33}=d_{33}(p)F_{(p,2)(22)}\left[\begin{array}{cc}(22)& (33)\\
(p-1,1)&(p,2) \end{array}  \right] =  \sqrt{\frac{p+1}{p-1}}\Bigl( 1 - \frac{\pi^2 p}{3m^2} + {\cal O}(m^{-3})\Bigr) 
\ee
\bea\label{A2}
C^{[(p+1,1)(p,2)(p,2)]22}_{22,33}&&=d_{33}(p)F_{(p,2)(22)}\left[\begin{array}{cc}(22)& (33)\\
(p+1,1)&(p,2) \end{array}  \right] \nonumber \\
&& =  -\sqrt{\frac{p-1}{p+1}}\Bigl( 1 + \frac{p\pi^2 }{3m^2} + {\cal O}(m^{-3})\Bigr) \, .
\eea
 The following  expressions are exact
\be
F_{(p,2)(22)}\left[\begin{array}{cc}(22)& (33)\\
(p-1,1)&(p,2) \end{array}  \right] = \frac{\Gamma\left(-\frac{3}{m+1}\right)\Gamma\left(\frac{2m+1-pm-p}{m+1}\right)
\Gamma\left(\frac{m+3}{m}\right)\Gamma\left(\frac{(p-1)(m+1)}{m}\right)}
{\Gamma\left(-\frac{1}{m+1}\right)\Gamma\left(-\frac{1+pm+p-2m}{m+1}\right)
\Gamma\left(\frac{m+1}{m}\right)\Gamma\left(\frac{pm+p-m+1}{m}\right)}
\ee
\be
F_{(p,2)(22)}\left[\begin{array}{cc}(22)& (33)\\
(p+1,1)&(p,2) \end{array}  \right] = -m
\frac{\Gamma\left(\frac{2m-1-p-pm}{m}\right)\Gamma\left(-\frac{3}{m+1}\right)
\Gamma\left(\frac{2m+1-p-pm}{m+1}\right)\Gamma\left(\frac{m+3}{m}\right)}
{\Gamma\left(-\frac{1}{m+1}\right)\Gamma\left(\frac{2m-1-p-pm}{m+1}\right)
\Gamma\left(\frac{2m+1-pm-p}{m}\right)\Gamma\left(\frac{1}{m}\right)}
\ee
Using these expressions we find 
\be \label{ratio_p} 
C^{[(p-1,1)(p,2)(p,2)]22}_{22,33}(C^{[(p+1,1)(p,2)(p,2)]22}_{22,33})^{-1}=-\frac{\sin\left(\frac{\pi(p+1)}{m}\right)}
{\sin\left(\frac{\pi(p-1)}{m}\right)}= -\frac{g_{p+1}}{g_{p-1}}
\ee
where $g_{p\pm 1}$ are the $g$-factors of the Cardy states $(p\pm 1,1)$: 
\be \label{gp}
g_{p\pm 1} = \left(\frac{8}{m(m+1)}\right)^{1/4}\frac{\sin\left(\frac{\pi(p\pm 1)}{m} \right)
\sin\left(\frac{\pi}{m+1} \right)}{\left(\sin\left(\frac{\pi}{m} \right) \sin\left(\frac{\pi}{m+1} \right) 
   \right)^{1/2}} \, . 
\ee
We further record the asymptotics for the following fusion matrices and OPE coefficients 
\be 
C_{13,13}^{[(p,2)(p,2)(p,2)]13} = d_{13}F_{(p,2)(13)}\left[\begin{array}{cc}(13)& (13)\\
(p,2)&(p,2) \end{array}  \right] \sim -\frac{4}{\sqrt{6}} \, , \enspace \mbox{ for any } p\ge 2\, ,
\ee
\bea 
&& F_{(p,2)(33)}\left[\begin{array}{cc}(33)& (33)\\
(p,2)&(p,2) \end{array}  \right]  = \frac{2}{p+1} -\frac{2}{3(p+1)m^2}\Bigl[ 
\pi^2p(p-1) + 6(\gamma + \psi(p-2)\nonumber \\
&&  + (p-2)\psi'(p-2))\Bigr] + {\cal O}\left(\frac{1}{m^3}\right)\, , \enspace p>2\, , 
\eea
\bea\label{333333}
C_{33,33}^{[(p,2)(p,2)(p,2)]33}&&=d_{33}(p)F_{(p,2)(33)}\left[\begin{array}{cc}(33)& (33)\\
(p,2)&(p,2) \end{array}  \right]  \nonumber \\
&& = \frac{2}{\sqrt{p^2-1}} - \frac{2\pi^2p^2}{3\sqrt{p^2-1}m^2} + {\cal O}\left(\frac{1}{m^3} \right) \, , \enspace 
p\ge 2\, ,
\eea
\be\label{333313}
C_{33,13}^{[(p,2)(p,2)(p,2)]33}=d_{13}(p)F_{(p,2)(33)}\left[\begin{array}{cc}(33)& (13)\\
(p,2)&(p,2) \end{array}  \right]  = \frac{4}{\sqrt{6}m} + {\cal O}\left(\frac{1}{m^2} \right)\, \enspace
p\ge 2\, ,
\ee
\be\label{333331}
C_{33,33}^{[(p,2)(p,2)(p,2)]31}=d_{31}(p)F_{(p,2)(33)}\left[\begin{array}{cc}(33)& (31)\\
(p,2)&(p,2) \end{array}  \right]  = \frac{4}{\sqrt{2(p^2-1)}m} + {\cal O}\left(\frac{1}{m^2} \right)\, \enspace
p\ge 2\, ,
\ee
\be\label{331331}
C_{33,13}^{[(p,2)(p,2)(p,2)]31}=\frac{d_{33}(p)d_{13}(p)}{d_{31}(p)}
F_{(p,2)(31)}\left[\begin{array}{cc}(33)& (13)\\
(p,2)&(p,2) \end{array}  \right]  = \frac{1}{\sqrt{3}} + {\cal O}\left(\frac{1}{m} \right)\, \enspace
p\ge 2\, .
\ee
\section{Conformal blocks}
\setcounter{equation}{0}

In this appendix $\phi_{i}$, $i=(i_1,i_2)\in K$ denote the Virasoro algebra chiral fields corresponding 
to the irreducible representations ${\cal H}_{p}$. Such a representation is obtained from the 
Virasoro Verma module built on descendants $L_{-n_1}L_{-n_2}\dots L_{-n_m}|h_{i}\rangle$ by taking 
the quotient with respect to singular vectors and their descendants. We have the usual $L_{0}$ 
grading: ${\cal H}_{p}=\oplus_{n=0}^{\infty}{\cal H}_{p}^{(n)}$ so that $L_{0}$ restricted to ${\cal H}_{p}^{(n)}$ 
equals $h_{p} + n$. 
 
A conformal block is formally defined by means of the expansion 
\be
{\cal F}_{ij;kl}^{p}(\eta) =  \sum_{K,K'}\eta^{h_{p}-h_{k}-h_{l}+|K|}\langle \phi_{i}|\phi_{j}(1)|\phi_{p}, K\rangle 
Q^{-1}_{K,K'}(p)\langle \phi_{p}, K'|\phi_{k}(1)|\phi_{l}\rangle  
\ee
where the indices $K$, $K'$ label the elements of a basis $|\phi_{p},K\rangle$
  in ${\cal H}_{p}^{(n)}$ with  $|K|=|K'|=n$. We can assume that the vectors $|\phi_{p},K\rangle$
  are linear combinations of vectors of the form $L_{-n_1}L_{-n_2}\dots L_{-n_m}|h_{p}\rangle$ 
  with $|K|=n_1+\dots + n_{m}$. The matrix $Q^{-1}_{K,K'}(p)$ is the inverse matrix to 
\be 
Q_{K,K'}(p) = \langle \phi_{p}, K|\phi_{p},K'\rangle  \, .
\ee
The matrix elements are defined as 
\be
\langle \phi_{i}|\phi_{j}(\eta) |\phi_{k}\rangle = {\cal N}_{jk}^{i} \eta^{h_{i}-h_{j}-h_{k}}\, , 
\ee
and the  operators $L_{n}$ act as  
\be
[L_{n}, \phi_{i}(\eta)] = {\cal L}_{n}^{h_{i}} \phi_{i}(\eta)\, 
\ee
where 
\be
{\cal L}_{n} ^{h_{i}}= \eta^{n+1}\partial_{\eta} + (n+1)h_{i}\eta^{n}   \, . 
\ee

We will use the following general formulas   
\be\label{num1}
\langle \phi_{i}|\phi_{j}(1)L_{-k_n}\dots L_{-k_{1}}|\phi_{p} \rangle = \prod_{i=1}^{n}
(h_p-h_i + k_{i}h_{j} + \sum_{s<i}k_{s}) \, , 
 \ee 
 \be\label{num2}
 \langle \phi_{p}|L_{k_{1}}\dots L_{k_{n}}\phi_{k}(1)|\phi_{l}\rangle 
 =\prod_{i=1}^{n}(h_p - h_l + k_{i}h_{k} + \sum_{s<i}k_{s}) \, . 
 \ee

The minimal model conformal blocks satisfy the following transformation rules 
\be\label{CT1}
{\cal F}_{ij;kl}^{p}\left(1-\eta \right)= \sum_{q}F_{pq}\left[\begin{array}{cc}j& k\\
i&l \end{array}  \right] {\cal F}_{il;jk}^{q}(\eta)\, ,
 \ee
\be\label{CT2}
{\cal F}_{ij;kl}^{p}\left(\frac{1}{\eta} \right)= \eta^{k_{k}+h_{l}+h_{j}-h_{i}} \sum_{q}B_{pq}^{(\pm)}\left[\begin{array}{cc}j& k\\
i&l \end{array}  \right] {\cal F}_{ik;jl}^{q}(\eta)
\ee
with 
\be\label{CT3}
B_{pq}^{(\pm)} \left[\begin{array}{cc}j& k\\
i&l \end{array}  \right] =F_{pq}\left[\begin{array}{cc}j& l\\
i&k\end{array}  \right] e^{\pm i\pi (h_{i}+h_{l}-h_{p}-h_{q})}\, . 
\ee
\subsection{Conformal blocks with $(2,2)$ fields}

Consider  a conformal block ${\cal F}_{22,31;22,13}^{22}(\eta)$. As $m\to \infty$ the weight 
$h_{22}$ goes to zero and the intermediate channel $\phi_{22}$ develops a zero norm vector 
$L_{-1}|\phi_{22}\rangle$. Thus the leading asymptotics should come from this singular vector and its 
descendants.  We  find from (\ref{num1}), (\ref{num2}) 
\bea\label{num3}
\langle \phi_{22}|\phi_{31}(1)L_{-k_n}\dots L_{-k_{1}}|\phi_{22} \rangle &=& 
\prod_{i=1}^{n}(\sum_{s\le i}k_{s}) + {\cal O}\left(\frac{1}{m}\right)\, ,  \nonumber \\
\langle \phi_{22}|L_{k_{1}}\dots L_{k_{n}}\phi_{22}(1)|\phi_{13}\rangle 
 &=&\prod_{i=1}^{n}(-1+ \sum_{s<i}k_{s})+ {\cal O}\left(\frac{1}{m}\right) \, . 
\eea
 Since any descendant of $L_{-1}|\phi_{22}\rangle$ is a linear combination of 
  vectors $L_{-k_n}\dots L_{-k_{2}}L_{-1}|\phi_{22}\rangle $ we see from (\ref{num3}) that the 
  leading contribution comes from $L_{-1}|\phi_{22}\rangle$ itself. Thus 
  \be
  {\cal F}_{22,31;22,13}^{22}( \eta) =  - \frac{h_{31}h_{13}}{2h_{22}}+ {\cal O}(m)= -\frac{2}{3}m^{2} 
+ {\cal O}(m) \, . 
  \ee
Analogously we obtain 
\be
{\cal F}_{22,31;22,31}^{22}( \eta ) =  - \frac{(h_{31})^2}{2h_{22}}+{\cal O}(m)=  -\frac{2}{3}m^{2}
+ {\cal O}(m) \, . 
\ee
We next take up the ${\cal F}_{22,31;22,31}^{42}( \eta )$ conformal block. We have 
\bea
\langle \phi_{42}|L_{k_{1}}\dots L_{k_{n}}\phi_{22}(1)|\phi_{31}\rangle 
 &=&(h_{42}-h_{31} + k_1 h_{22})\prod_{i=2}^{n}(h_{42}-h_{31} + k_{i}h_{22} + \sum_{s<i}k_{s})+ {\cal O}\left(\frac{1}{m}\right) \nonumber \\ &=& {\cal O}\left(\frac{1}{m}\right) \, , 
\eea
and thus 
\be
{\cal F}_{22,31;22,31}^{42}( \eta ) = 1 + {\cal O}\left(\frac{1}{m}\right) \, . 
\ee

The leading order contribution to $\tilde {\cal F}_{22,31;22,33}^{22}( \eta )$ comes from the asymptotic 
singular vector $L_{-1}|\phi_{22}\rangle$ and its descendants. 
Let $L_{K}=L_{k_1}L_{k_2}\dots L_{k_n} $ with $ k_{i}\ge 1$. We have 
\be
\langle \phi_{22} | L_{1}L_{K}\phi_{22}(\eta)L_{-1}|\phi_{33}\rangle =h_{33}(1-h_{33}+2h_{22}) {\cal L}_{K}^{h_{22}}\eta^{-h_{33}} + \langle \phi_{22} | L_{1}\phi_{22}(\eta)[L_{K},L_{-1}]|\phi_{33}\rangle
\ee
where 
\be
{\cal L}_{K}^{h_{22}}= {\cal L}_{k_{n}} ^{h_{22}}\dots {\cal L}_{k_1}^{h_{22}} \, .
\ee
Since
\be
[L_K,L_{-1}]=\sum_{|K'|=|K|-1}\alpha_{K'}L_{K'} 
\ee
we have
\be
\langle \phi_{22} | L_{1}\phi_{22}(\eta)[L_{K},L_{-1}]|\phi_{33}\rangle\sim h_{33}{\cal L}_{1} \eta^{-h_{33}} = h_{33}(-h_{33}+2h_{22})\eta^{1-h_{33}} \, . 
\ee

Since $h_{22}\sim m^{-2}$, $h_{33}\sim m^{-2}$ we conclude that
\be
\langle \phi_{22} | L_{1}L_{K}\phi_{22}(\eta)L_{-1}|\phi_{33}\rangle ={\cal O}\left( \frac{1}{m^4}\right) \, 
\ee
when $|K|=k_1+\dots + k_n> 0$. Thus the leading contribution comes from  $L_{-1}|\phi_{22}\rangle$ and can be readily evaluated:
\be
\tilde {\cal F}_{22,31;22,33}^{22}( \eta ) = \frac{h_{31}(2h_{22}h_{33} + h_{33}(1-h_{33}))}{2h_{22}}\eta^{-h_{33}} + {\cal O}\left(  \frac{1}{m^2}\right) = 
\frac{4}{3} + {\cal O}\left(  \frac{1}{m}\right) \, . 
\ee

Similarly we have 
\bea
 \langle \phi_{42}| L_{K}\phi_{22}(\eta)L_{-1}|\phi_{33}\rangle&&  = (h_{22}+h_{33}-h_{42}){\cal L}_{K}\eta^{h_{42}-1-h_{22}-h_{33}} \nonumber \\
&&   + \langle \phi_{42}| \phi_{22}(\eta)[L_{K},L_{-1}]|\phi_{33}\rangle  
\eea
and thus 
\be
 \langle \phi_{42}| L_{K}\phi_{22}(\eta)L_{-1}|\phi_{33}\rangle = {\cal O}\left( \frac{1}{m}\right)
 \ee
if $|K|>0$. This implies 
\be
\tilde {\cal F}_{22,31;22,33}^{42} = -1 +  {\cal O}\left( \frac{1}{m}\right)\, . 
\ee

\subsection{Conformal blocks for $1/m$ corrections}
In this appendix we  derive the leading asymptotics of ${\cal F}_{33,33;33,13}^{33}(\eta)$ and ${\cal F}_{33,33;33,13}^{31}(\eta)$. 
We will use a method different from the method of section B.1. Note that at $m=\infty$ the conformal dimensions of $h_{33}$, $h_{13}$, $h_{31}$ become integers 
so that the leading asymptotics of the conformal blocks at hand should be given by rational functions. The behaviour  of these rational functions at $\eta=0,1,\infty$ 
can be obtained using (\ref{CT1})-(\ref{CT3}). 
We record the following leading asymptotics of  the relevant    fusion and braiding matrices 
\be
F_{(33)(33)}\left[\begin{array}{cc}(33)& (13)\\
(33)&(33) \end{array}  \right]  \sim \frac{1}{2}\, , \quad 
F_{(33)(31)}\left[\begin{array}{cc}(33)& (13)\\
(33)&(33) \end{array}  \right]  \sim \frac{1}{3}\, ,   
\ee 
\be
F_{(33)(35)}\left[\begin{array}{cc}(33)& (13)\\
(33)&(33) \end{array}  \right]  \sim \frac{5}{12}  \, , \quad F_{(31)(33)}\left[\begin{array}{cc}(33)& (13)\\
(33)&(33) \end{array}  \right]  \sim 1 \, , 
\ee
\be
F_{(31)(31)}\left[\begin{array}{cc}(33)& (13)\\
(33)&(33) \end{array}  \right]  \sim \frac{1}{3}  \, , \quad F_{(31)(35)}\left[\begin{array}{cc}(33)& (13)\\
(33)&(33) \end{array}  \right]  \sim -\frac{5}{6} \, , 
\ee
\be
B_{(33)(33)}^{(\pm)}\left[\begin{array}{cc}(33)& (33)\\
(33)&(13) \end{array}  \right]  \sim- \frac{1}{2}\, , \quad 
B_{(33)(31)}^{(\pm)}\left[\begin{array}{cc}(33)& (33)\\
(33)&(13) \end{array}  \right]  \sim \frac{1}{3}\, ,   
\ee 
\be
B_{(33)(35)}^{(\pm)}\left[\begin{array}{cc}(33)& (33)\\
(33)&(13) \end{array}  \right]  \sim  \frac{5}{12}\, , \quad 
B_{(31)(33)}^{(\pm)}\left[\begin{array}{cc}(33)& (33)\\
(33)&(13) \end{array}  \right]  \sim  1 \, ,   
\ee 
\be
B_{(31)(31)}^{(\pm)}\left[\begin{array}{cc}(33)& (33)\\
(33)&(13) \end{array}  \right]  \sim  -\frac{1}{3}\, , \quad 
B_{(31)(35)}^{(\pm)}\left[\begin{array}{cc}(33)& (33)\\
(33)&(13) \end{array}  \right]  \sim  \frac{5}{6} \, .   
\ee 

We have
\be \label{F0}
{\cal F}_{33,33;33,13}^{33}(\eta) = \eta^{-h_{13}}   +\frac{2h_{33}-h_{13}}{2}\eta^{1-h_{13}}  + \dots \sim \frac{1}{\eta} -\frac{1}{2}   \, , \quad \eta \sim 0 \, .  
\ee
Using (\ref{CT1})-(\ref{CT3})  we also obtain 
\bea
{\cal F}_{33,33;33,13}^{33}(\eta) && = \frac{5}{12}{\cal F}_{33,13;33,33}^{35}(1-\eta) + \frac{1}{3} {\cal F}_{33,13;33,33}^{31}(1-\eta)\nonumber \\
&& + \frac{1}{2}{\cal F}_{33,13;33,33}^{33}(1-\eta) + {\cal O}\left(\frac{1}{m}\right) \, , 
\eea 
\bea
{\cal F}_{33,33;33,13}^{(33)}(\eta) && = \eta^{-h_{13}-h_{33}}\Bigl[ \frac{5}{12}{\cal F}_{33,33;33,13}^{35}\left(\frac{1}{\eta}\right) + \frac{1}{3} 
{\cal F}_{33,33;33,13}^{31}\left(\frac{1}{\eta}\right) \nonumber \\
&& - \frac{1}{2}{\cal F}_{33,33;33,13}^{33}\left(\frac{1}{\eta}\right) \Bigr]+ {\cal O}\left(\frac{1}{m}\right) \, .
\eea

Noting that  for $\eta \sim 1$ we have an expansion 
\be 
{\cal F}_{33,13;33,33}^{33}(1-\eta) = (1-\eta)^{-h_{33}} + \frac{h_{13}}{2}(1-\eta)^{1-h_{33}} + {\cal O}\left((1-\eta)^{2-h_{33}}\right)
\ee
we find from the above 
\bea\label{F1}
{\cal F}_{33,33;33,13}^{33}(\eta)& \sim & \frac{1}{2} + (1-\eta) + \dots  \, , \enspace  \eta \to 1  \, , \nonumber \\
{\cal F}_{33,33;33,13}^{33}(\eta)& \sim &- \frac{1}{2} + \frac{1}{\eta}  + \dots  \, , \enspace  \eta \to \infty 
\eea
up to terms suppressed by $1/m$. It follows from (\ref{F0}), (\ref{F1}) that 
\be 
{\cal F}_{33,33;33,13}^{33}(\eta) = \frac{1}{\eta}  - \frac{1}{2} +  {\cal O}\left(\frac{1}{m}\right) \, . 
\ee

We further find 
\bea
{\cal F}_{33,33;33,13}^{31}(\eta) && = {\cal F}_{33,13;33,33}^{33}(1-\eta) + \frac{1}{3}{\cal F}_{33,13;33,33}^{31}(1-\eta)\nonumber \\
&&   -\frac{5}{6}{\cal F}_{33,13;33,33}^{35}(1-\eta) + {\cal O}\left(\frac{1}{m}\right) \, , 
\eea 
\bea
{\cal F}_{33,33;33,13}^{31}(\eta) && = \eta^{-h_{33}-h_{13}}\Bigl[ {\cal F}_{33,33;33,13)}^{33}\left(\frac{1}{\eta}\right) - \frac{1}{3}{\cal F}_{33,33;33,13}^{31}\left(\frac{1}{\eta}\right) \nonumber \\
&&   +\frac{5}{6}{\cal F}_{33,33;33,13}^{35}\left(\frac{1}{\eta}\right)  \Bigr]  + {\cal O}\left(\frac{1}{m}\right) \, .
\eea 
From this we find that ${\cal F}_{33,33;33,13}^{31}(\eta) \sim 1 $ through the first two orders near $\eta=0,1,\infty$ and thus 
\be
{\cal F}_{33,33;33,13}^{31}(\eta) =1 + {\cal O}\left(\frac{1}{m}\right) \, . 
\ee


\begin{thebibliography}{99}
\bibitem{Gaiotto} D. Gaiotto, {\it Domain walls for two-dimensional renormalization group flows}, 
arXiv:1201.0767.


\bibitem{GuidaMagnoli} R. Guida and N. Magnoli, 
{\it All order I.R. finite expansion for short distance behavior of massless theories 
perturbed by a relevant operator}, 
Nucl. Phys. {\bf B471} (1996) 361-388; arXiv: hep-th/9511209.

\bibitem{AleshaZam} Al. B. Zamolodchikov, {\it Two-point correlation function in scaling Lee-Yang model}, Nucl. Phys. {\bf B348} (1991) 619.

\bibitem{Litvinovetal} A.A.Belavin, V.A.Belavin, A.V.Litvinov, Y.P.Pugai and  Al.B.Zamolodchikov, 
{\it On correlation functions in the perturbed minimal models $M(2,2n+1)$}, 
Nucl. Phys. {\bf B676} (2004) 587-614; arXiv: 1. arXiv:hep-th/0309137.

\bibitem{Zamolodchikov_pert} A. B. Zamolodchikov, {\it Renormalization group and 
perturbation theory about fixed points in two-dimensional field theory}, 
Sov. J. Nucl. Phys. {\bf 46} (1987) 1090.



\bibitem{PetkovaZuber} V. B. Petkova and J. B. Zuber, 
{\it Generalised twisted partition functions}, Phys. Lett. {\bf B504} (2001) 157Ð164; arXiv:hep-th/0011021.

\bibitem{Bachasetal} C. Bachas, J. de Boer, R. Dijkgraaf, and H. Ooguri, 
{\it Permeable conformal walls andholography},  JHEP {\bf 06} (2002) 027; arXiv: hep-th/0111210.

\bibitem{GrahamWatts}K. Graham and G.M.T. Watts, {\it  Defect Lines and Boundary Flows}, JHEP 0404 (2004) 019; arXiv:arXiv:hep-th/0306167.

\bibitem{Frohlichetal} J. Froehlich, J. Fuchs, I. Runkel, and C. Schweigert, {\it Kramers-Wannier duality fromconformal defects}, Phys. Rev. Lett. {\bf 93} (2004) 070601; arXiv: cond-mat/0404051.
 
 \bibitem{BachasGaberdiel} C. Bachas and M. Gaberdiel, {\it Loop operators and the Kondo problem}, JHEP {\bf 11}(2004) 065;arXiv: hep-th/0411067.
 
  \bibitem{BR} I. Brunner and D. Roggenkamp, {\it Defects and bulk perturbations of boundary 
  Landau-Ginzburg orbifolds}, JHEP {\bf 0804} (2008) 00; arXiv:0712.0188.
  
  \bibitem{FQ} S. Fredenhagen and T. Quella, {\it Generalised permutation branes},  JHEP {\bf 0511} (2005) 004; 
  arXiv:hep-th/0509153. 
  
  
\bibitem{GRW} K. Graham, I. Runkel and G.M.T. Watts, {\it Minimal model boundary flows and $c=1$ CFT},  
Nucl.cPhys. {\bf B608} (2001) 527-556;  arXiv:hep-th/0101187. 

\bibitem{Runkel}  I. Runkel, {\it Boundary structure constants for the 
A-series Virasoro minimal models}, Nucl.Phys. B549 (1999) 563-578; arXiv:hep-th/9811178.

\bibitem{RRS} A. Recknagel, D. Roggenkamp and V. Schomerus, 
{\it On relevant boundary perturbations of unitary minimal models}, 
Nucl. Phys. {\bf B 588} (2000) 552-564; arXiv:hep-th/0003110.

\bibitem{AL} I. Affleck and A. W. W. Ludwig, {Universal noninteger `ground state degeneracy' in critical quantum 
systems}, Phys. Rev. Lett. {\bf 67} (1991) 161.

\bibitem{withCor} A. Konechny and C. Schmidt-Colinet, unpublished.








\end{thebibliography}
\end{document}